\documentstyle[aps,prd,epsfig,floats,preprint,tighten]{revtex}
\newcommand{\Pom}{I$\!$P}
\newcommand{\ffig}[4]{\begin{figure}[htb]\vfill\begin{center}
\mbox{\epsfig{figure=#1,height=#2}}\caption{#3}\label{#4}
\end{center}\vfill\end{figure}}
\begin{document}
\preprint{TSL/ISV-95-0125, DESY-95-163}
\title{Soft Colour Interactions as the Origin of Rapidity Gaps in DIS}
 
\author{A.~Edin$^1$, G.~Ingelman$^{1,2}$, J.~Rathsman$^1$}
\address{ $^{1}$Dept. of Radiation Sciences, Uppsala University, 
 Box 535, S-751 21 Uppsala, Sweden\\
 $^{2}$Deutsches Elektronen-Synchrotron DESY, Notkestrasse 85, D-22603 Hamburg,
 Germany}

\maketitle

\begin{abstract}
We introduce soft colour interactions as a novel mechanism to 
understand the observed events with large rapidity gaps in $ep$ collisions at
HERA. Colour exchanges between produced partons and colour-charges in the
proton  remnant modifies the colour structure for hadronization, such that
colour  singlet systems may appear well separated in rapidity. Our explicit
model show  characteristics of diffractive scattering, although no explicit
pomeron  dynamics have been introduced, and for non-gap events an increased
forward  energy flow gives agreement with data. 
\end{abstract}

In deep inelastic scattering (DIS) at the HERA $ep$ collider a relatively large
fraction ($\sim 10\%$) of events have been observed \cite{ZEUS,H1} to have 
a rapidity gap, i.e.\ no particles or energy in a large rapidity region close 
to the proton beam direction. These events can be 
interpreted in terms of hard scattering on a pomeron (\Pom ) \cite{IS}, 
a colour singlet object exchanged in a Regge description of diffractive 
interactions. In particular, models (see e.g.\ \cite{GI,Landshoff})
with a factorization of a pomeron flux and a pomeron-particle hard scattering 
cross section, using parton density distributions in the pomeron,  
can describe the salient features of the observations. 
Nevertheless, there is no 
satisfactory understanding of the pomeron and its interaction mechanism. 

A main conceptual problem is whether the scattering is on a preformed colour 
singlet object (\Pom ). 
This need not be the case since soft interactions could take 
place with the proton both before and after the hard scattering
in such a way that a colour singlet system is formed leaving a well separated
forward proton (or small-mass system). In the spirit of the latter scenario
we have suggested \cite{EIR} a novel way to interprete the rapidity gap 
phenomenon, without using the concept of a pomeron.
 
Here we present our model, which is based on the new
hypothesis that soft colour interactions change the hadronization such that
rapidity gaps occur in the final state. The starting point is the normal DIS
parton interactions, with perturbative QCD (pQCD) corrections,  giving a state
of partons to be hadronized. By assuming that these partons undergo
non-perturbative soft colour interactions the colour structure will  change
such that when normal hadronization models are applied rapidity gaps  may
arise. 

At small Bjorken-$x$ ($10^{-4}-10^{-2}$), where the rapidity gap events are
observed,  the boson-gluon-fusion (BGF) process $\gamma g\to q\bar{q}$
(cf.~Fig.~1) constitutes a substantial part of the cross-section.
This process is calculable in first order
QCD, with the conventional requirement $m_{ij}^2>y_{cut}W^2$ on any pair $ij$
of partons to avoid soft and collinear divergences. Higher order pQCD emissions
can be taken  into account approximately through parton shower evolution   from
the final partons and the incoming one (as illustrated with one  emitted gluon
in Fig.~1). In the following non-perturbative hadronization  process one
usually considers the formation of colour singlet systems  (clusters, strings)
that subsequently break up into hadrons.  In the conventional Lund model
\cite{Lund} treatment, a BGF event gives  two separate strings from the $q$ and
$\bar{q}$ to the proton remnant spectator partons (Fig.~1a), thereby causing
particle production over the whole rapidity region in between.  This treatment
is used in the Monte Carlo LEPTO \cite{LEPTO}, which describes  
most features of HERA DIS events. 

\ffig{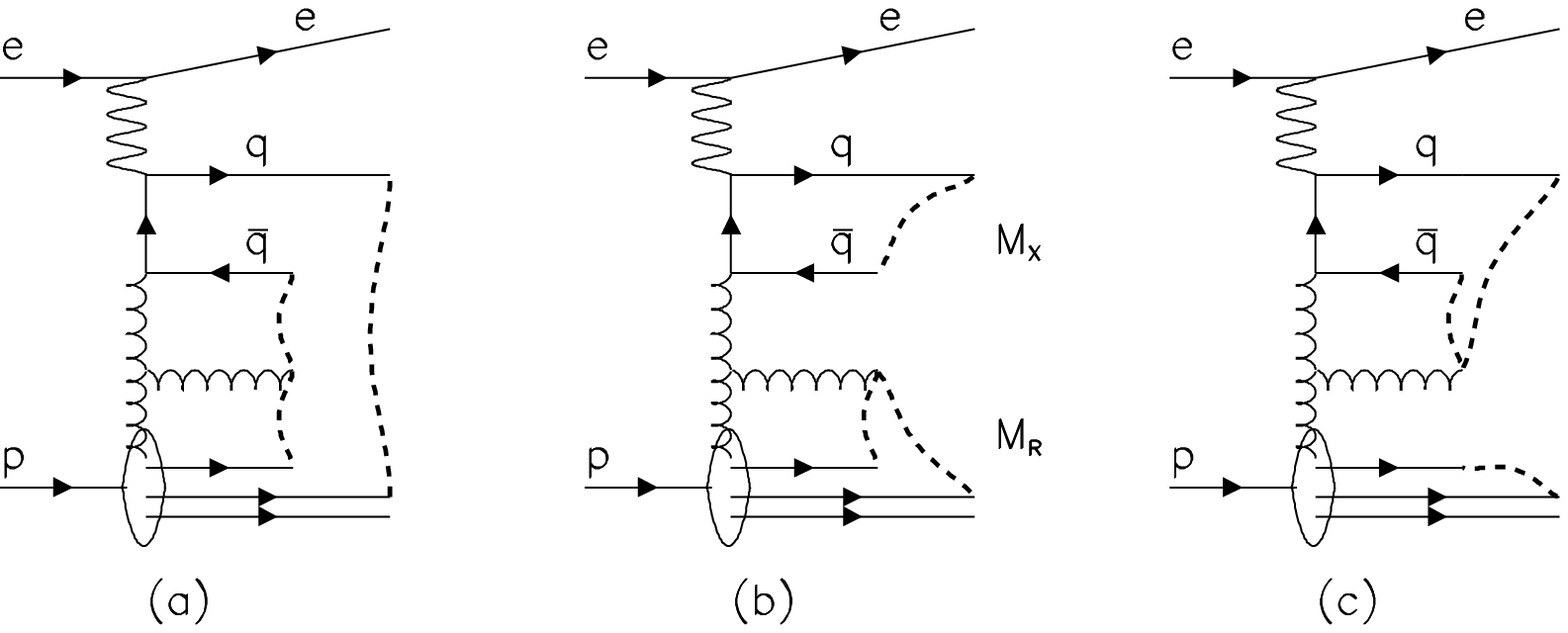}{50mm}{The string configuration in a DIS 
boson-gluon-fusion event: 
(a) conventional Lund string connection of partons, and 
(b,c) after reconnection due to soft colour interactions.}{fig1}

This conventional treatment assumes that the colour structure, i.e.\ the 
string topology, follows exactly the colour ordering from the
perturbative phase with no further 
alterations. Our main assumption here is that additional non-perturbative 
soft colour interactions (SCI) may occur. These have small momentum  transfers,
below the scale $Q_0^2$ defining the limit of pQCD, and do not  significantly
change the momenta from the perturbative phase.  However, SCI will change the
colour of the partons involved and thereby change the colour topology as
represented by the strings. Thus, we propose that the perturbatively produced
quarks and gluons can interact softly with the colour medium of the proton
as they propagate through it. This should be a natural part of the
processes in which `bare'  perturbative partons are `dressed'  into
non-pertubative quarks and gluons and the formation of the confining  colour
flux tube in between them. 

Lacking a proper understanding of such non-perturbative QCD processes, we 
construct a simple model to describe and simulate these interactions. All
partons from the hard interaction (electroweak $+$ pQCD) plus the  remaining
quarks in the proton remnant constitute a set of colour charges. Each pair of
charges can make a soft interaction changing only the colour and not the
momenta,  which may be viewed as soft non-perturbative gluon exchange.  As the
process is non-perturbative the exchange probability for a pair  cannot be
calculated so instead we describe it by a phenomenological   parameter $R$. 
The number of soft exchanges will vary event-by-event and change  the colour
topology of the events such that, in some cases, colour singlet subsystems
arise separated in rapidity. In the Lund model this corresponds to a modified
string stretching as illustrated in Figs.~1bc, where (b) can be seen as a
switch of anticolour between the antiquark and the diquark and (c) as a switch
of colour between the two quarks. This kind of colour switches between the
perturbatively produced partons and the partons in  the proton remnant are of
particular importance for the gap formation.

\ffig{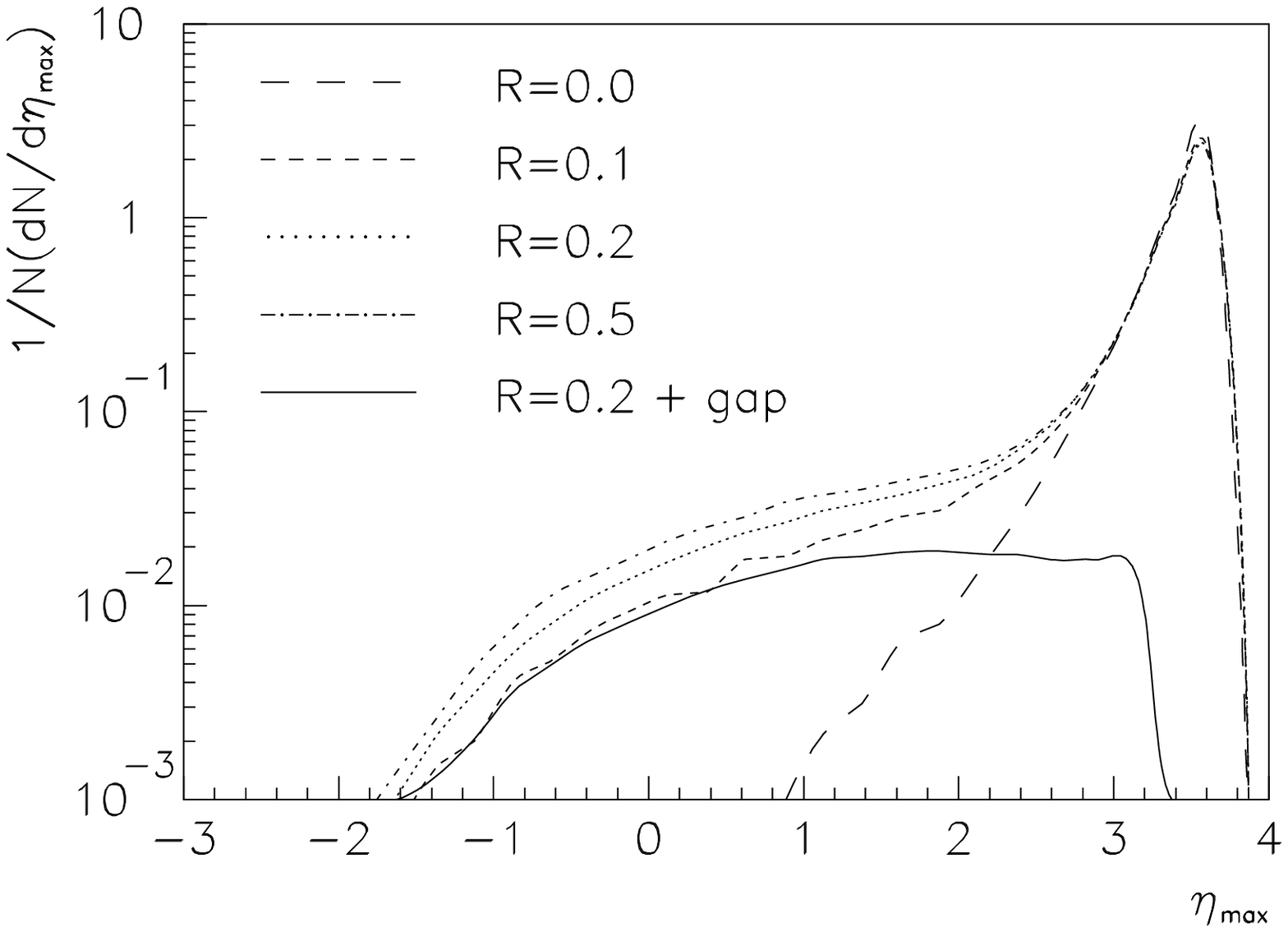}{70mm}{Distribution of maximum pseudorapidity 
($\eta_{max}$). Hadron level after colour reconnection with probability 
parameter $R$ for all events and those satisfying the `gap' definition 
(full line).}{fig2}

Rapidity gaps have been experimentally investigated \cite{ZEUS,H1} through the
observable $\eta_{max}$ giving, in each event, the maximum  pseudo-rapidity
where an energy deposition is observed. (With $\eta =-\ln{\tan{\theta/2}}$
and $\theta$ the angle relative to the proton beam so that $\eta > 0$ is the
proton hemisphere in the HERA lab frame.) Fig.~2 shows the distribution of
this quantity as obtained from our model simulations for $7.5<Q^2<70$ and
$0.03<y<0.7$, corresponding to the experimental conditions. In addition, the
limit $\eta_{max}<3.65$ in the H1 detector is taken into account. Clearly, the
introduction of soft colour interactions ($R>0$) have a large effect on the
$\eta_{\max}$ distribution. Still, our SCI model is not very  sensitive to the
exact value of the parameter $R$. In fact, increasing $R$ above $0.5$ does not
give an increased gap probability, but may actually decrease it depending on
the details of the colour exchanges in the model. This is intuitively
understandable, since once a  colour exchange with the spectator has occured
additional exchanges among the partons need not favour gaps and may even
reduce them. In the following we use $R=0.2$. This value may be seen as the
strong coupling $\alpha_s(0.5\, GeV)/\pi \approx 0.2$ at a momentum transfer 
representative for the region below the perturbative cutoff $Q_0^2\sim 1\:
GeV^2$.

One should note that the basic features of this distribution, the height of the
peak and the `plateau', is in reasonable agreement with the data \cite{ZEUS,H1}.
A direct comparison requires a detailed account of experimental conditions, 
such as acceptance and varying event vertex position. 
Selecting events with rapidity gaps similar to the H1 definition 
(i.e.\ no energy in $\eta_{max}<\eta<6.6$ where $\eta_{max}<3.2$) 
gives the full curve in Fig.~2, also in basic agreement with data \cite{H1}.

\ffig{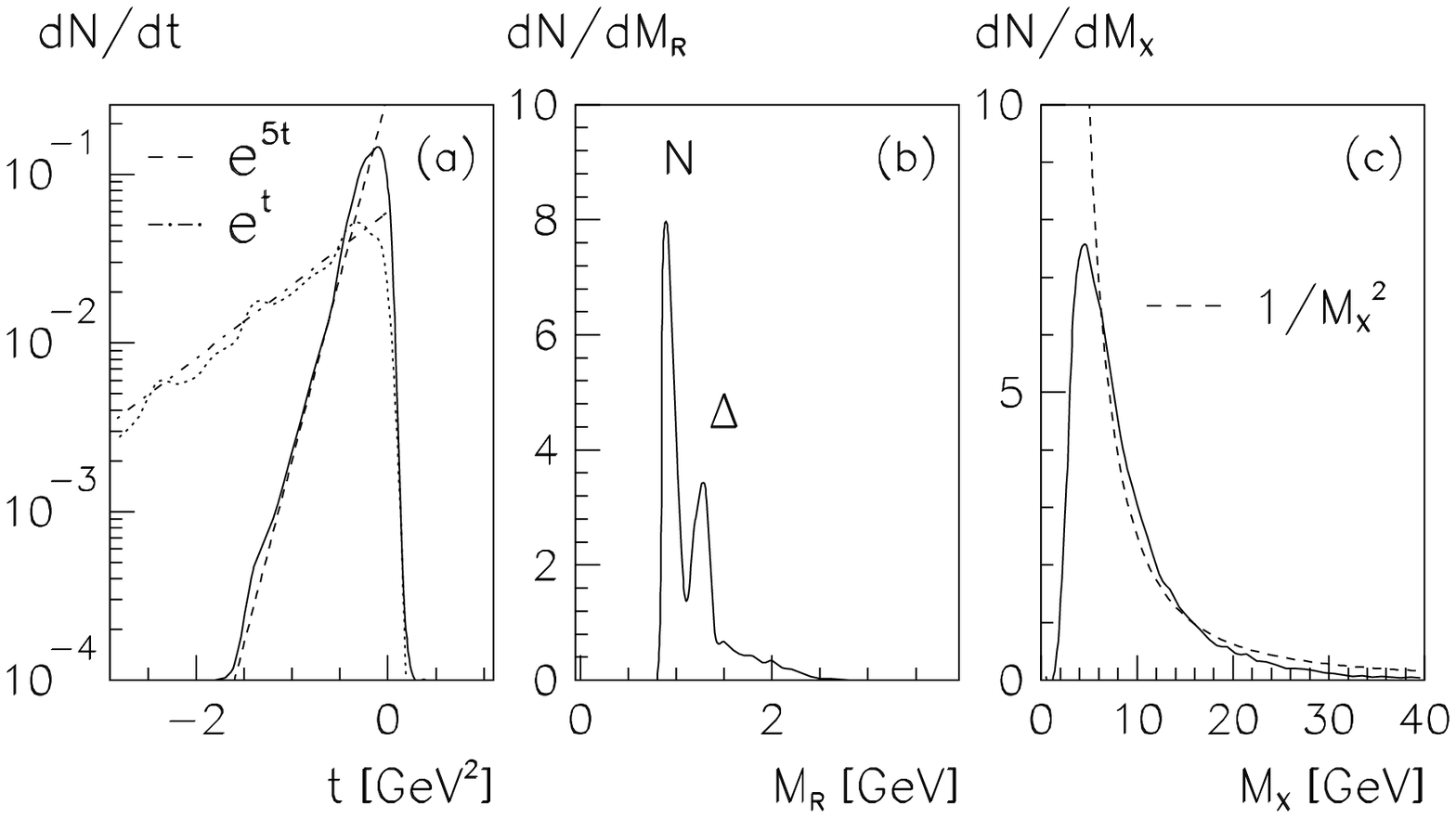}{70mm}{Distributions (in arbitrary units) for the selected 
rapidity gap events,  i.e.\ with no energy in $\eta_{max}<\eta<6.6$ and
$\eta_{max}<3.2$: (a) squared momentum transfer $t$ from incoming proton to
remnant system $R$, for two widths $\sigma =0.44$ (full) and 1 GeV (dotted)
of the gaussian 
primordial $k_\bot$ distribution compared with exponential slopes $1/\sigma^2$;
(b,c) invariant mass of the forward remnant system $M_R$
and the produced central system $M_X$ (cf.~Fig.~1)}{fig3}

Further features of our model are shown in Fig.~3, where the resulting 
distributions in momentum transfer $t=(p_p-p_R)^2$ and mass of the remainder 
system $R$ and
the produced system $X$ (cf.~Fig.~1) are displayed for the selected  gap
events.   Although the model makes no particular assumptions or requirements on
these quantities for the gap events, their distributions are similar to  what
is expected from diffractive models. This applies to the essentially
exponential $t$-dependence,  $1/M_X^2$ dependence and the $M_R$ system being
dominated by the proton. 

The $t$-dependence in our model is intimately connected to the assumed 
distribution of primordial transverse momentum $k_{\bot}$ of partons in the
proton, i.e.\ of the parton entering the hard scattering process.
This transverse momentum is balanced by the proton remnant and, since 
momentum transfers in SCI are neglected, it is essentially the $p_\bot$ 
of the forward $R$-system, i.e.\ $p_{R,\bot}^2=k_{\bot}^2$. 
Now, $t\approx -p_{R,\bot}^2$ in the case of interest, i.e.\ when the 
energy-momentum transfer from the beam proton to the $X$-system is small 
giving a very forward $R$-system.
The primordial $k_\bot$ represents the non-perturbative Fermi motion in the
proton and is therefore of the order $k_{\bot}\simeq 1 \: \mbox{fm}^{-1}$ or 
a few
hundred $MeV$ as estimated from the uncertainty principle. This gives the
width $\sigma$ of the Gaussian distribution
$\exp{(-k_{\bot}^2/\sigma^2)}dk_{\bot}^2$ which is normally assumed. 
Thus, one directly gets the exponential $t$-dependence $\exp{(t/\sigma^2)}dt$ 
with $\sigma^2=2\langle k_{\bot}^2 \rangle$ from the primordial 
$k_\bot^2$-distribution.  
As demonstrated in Fig.~3a, the input $k_\bot^2$-distribution is very well 
reproduced in the $t$-distribution of the Monte Carlo events.

The $R$-system is dominantly a single proton, as in a diffractive model based
on pomeron exchange. However, there is also a substantial amount of $\Delta$ 
which would correspond to pion exchange in Regge phenomenology. The detailed
composition of the $R$-system does in our case depend on the model 
\cite{JETSET} used for hadronizing a small-mass string system. 
This model is not constructed to give a detailed account of quantum numbers
and masses in the resonance region, but rather a reasonable mean behaviour.  
This leaves some room for modifications that may change the detailed outcome,
but still it should be possible to find distinguishing features in the 
$R$-system in comparison with Regge-based models. 

The $1/M_X^2$ behaviour is explained by the $1/s_{q\bar{q}}$ dependence of the
BGF matrix elements, but is distorted at large $M_X$ by requiring the gap to
extend into the central rapidity region. Kinematically, larger $M_X$ means a
reduced gap  and is therefore disfavored by the gap condition.

Thus, with a gap definition suitable for selecting diffractive interactions,
our model shows the same general  behaviour as models based on pomeron and
other Regge exchanges. One must however keep in mind that the experimental
conditions for  selecting gap events, requiring a large gap that extends very
forward  in rapidity, gives a kinematical bias against large values of  $t$ and
$M_R$.  Given the different input concerning the formation of the forward
system in  our model and in pomeron models, it seems likely that observable
differences should occur when varying the gap definition or observing the
forward-moving $R$-system.

\ffig{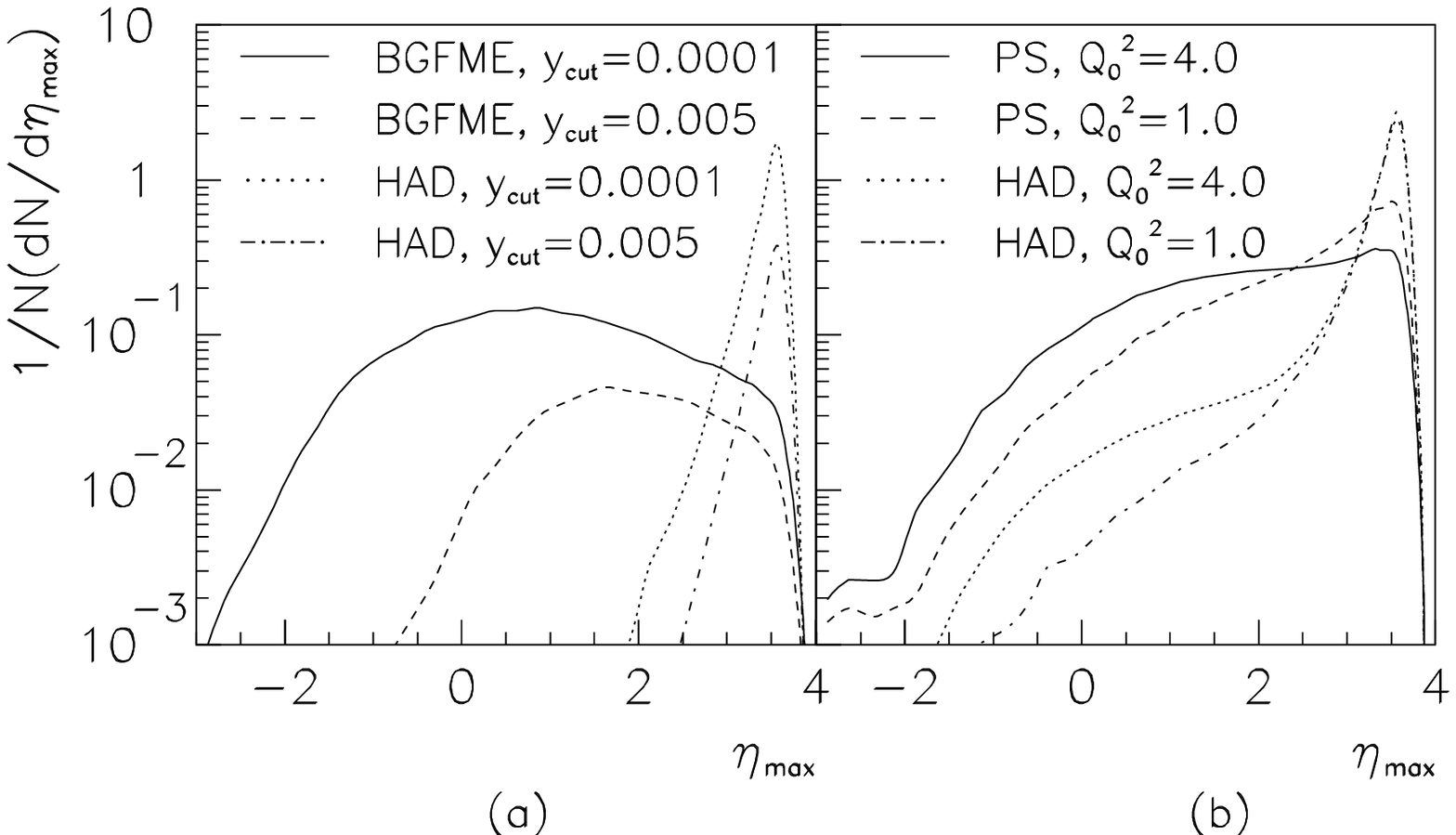}{60mm}{Distribution of maximum pseudorapidity $\eta_{max}$.
(a) Partons from boson-gluon-fusion matrix element (BGFME)  with cut-off
$y_{cut}$, and hadrons (HAD) after parton showers and string hadronization
without SCI.  
(b) Parton level (PS) and hadron level (HAD), after SCI, 
for different initial state  parton shower cut-off values $Q_0^2$.}{fig4}

The probability for gaps from SCI depends on the parton state used as starting
configuration. Therefore, we must consider the influences from variations and 
uncertainties related to both the matrix elements and the higher order
parton shower emissions. 
The maximum rapidity parton from the BGF matrix element  (Fig.~4a) can be
central or even in the electron beam hemisphere depending  on the phase space
allowed by $y_{cut}$. For $y_{cut}=0.005$, which has been shown to be
theoretically sound \cite{scale}, about 10 \% are BGF events. The small
$y_{cut}=0.0001$ results in an automatic adjustment \cite{LEPTO} of the 
cut-off such
that the total (Born) cross section is saturated with $2+1$-jet  events, giving
$\sim 50\%$ BGF events. 
With the smaller $y_{cut}$ the partons can 
emerge with a large rapidity gap relative to the spectator partons 
(at large $\eta$ outside the scale of the figure and experimentally lost in the 
beam pipe). 
This gives a larger potential to produce gaps in the final state and 
$y_{cut}=0.0001$ is therefore used as default in our model. 
Fig.~4a also demonstrates that normal parton showers and string hadronization 
give so large effects that those potential gaps does not survive in the 
absence of SCI. 

Since parton showers and hadronization give so large effects 
the exact gap probability will depend on details of these. 
In particular, the cut-off $Q_0^2$ for the  parton shower is rather important.
Chosing a value close the hadronic mass scale $\sim 1\: GeV^2$ tends to produce
too much radiation at large rapidity  such that the gaps are partly destroyed
(Fig.~4b).  A value of $4 \: GeV^2$ for the limit  of pQCD, as in many parton
density parametrizations, reduces such emissions  and thereby larger gaps can
arise after SCI. This has been taken as  the default value in our model.  

In this context one should note that the conventional leading $\log{Q^2}$ 
evolution need not be correct when applied to the treatment of exclusive parton
final states in a parton shower. It is derived \cite{AP} for not-too-small $x$
and only for the inclusive case, i.e.\ for the evolution of the structure
function. It therefore  sums over all emissions such that important
cancellations can be exploited. It is not clear whether this formalism is fully
applicable also to exclusive final states. It seems likely that it gives the
correct mean behaviour, but it may not properly estimate the fluctuations that
can occur in the emission chain. Some events may therefore have less parton
radiation than estimated in this way and these would favour the occurence of
rapidity gaps. 

\ffig{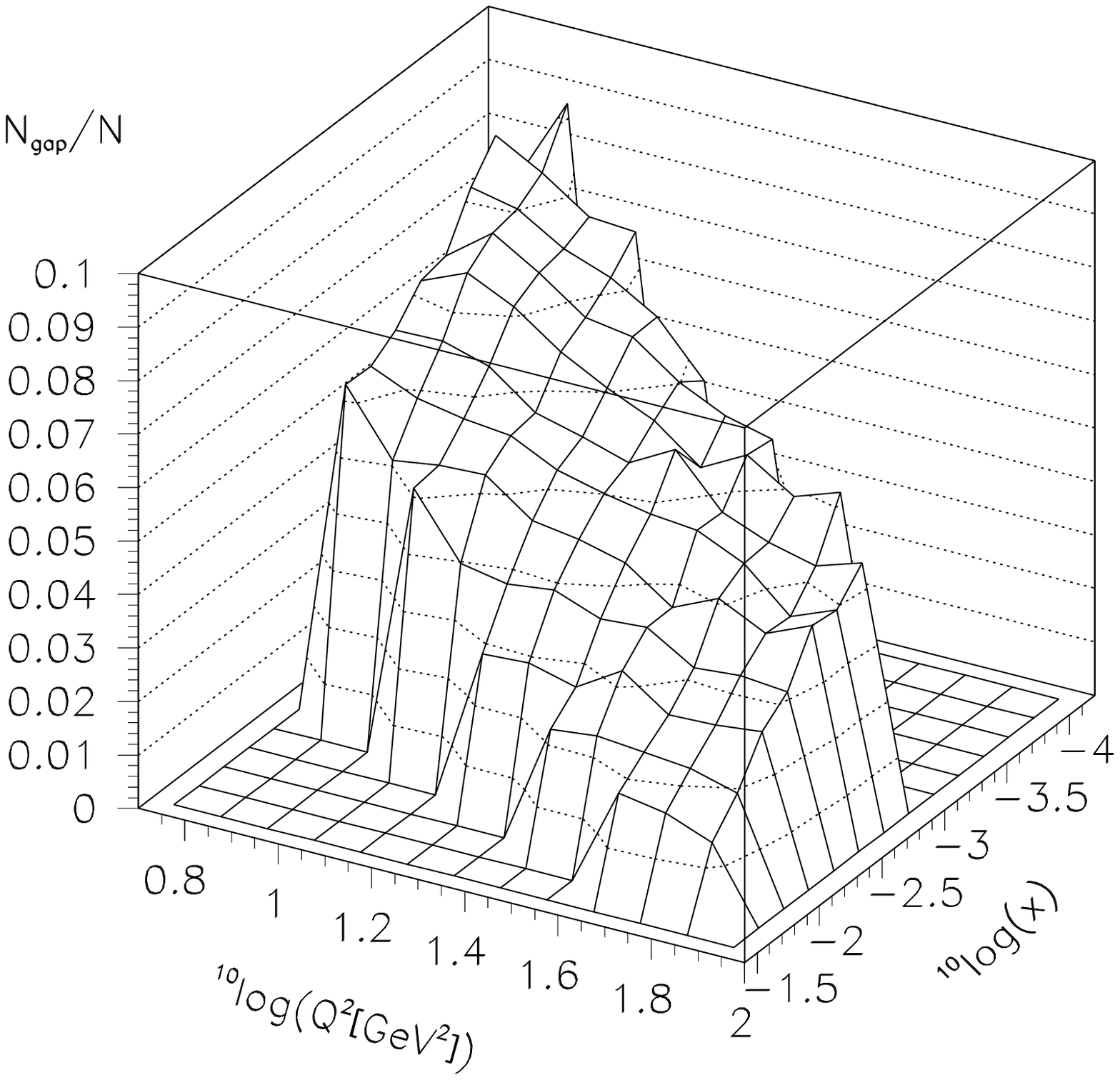}{90mm}{Ratio of rapidity gap events, 
i.e.\ with no energy in $\eta_{max}<\eta<6.6$ and $\eta_{max}<3.2$,
to all DIS events as function of $x$ and $Q^2$.}{fig5}

Since the soft colour interactions in our model are factorized with respect to
the hard interaction, one may expect that the rate of gap events is
essentially  independent of the DIS kinematical variables $x$ and $Q^2$. This
is, however, not quite the case as shown in Fig.~5. The variation in $x$, at
fixed $Q^2$ is significant. A smaller part of the effect is from the increase
of gluon induced hard scatterings as $x$ decreases. However, most of it is a
purely kinematic effect due to the gap definition. With increasing $x$, i.e.\
harder incoming parton, the hard scattering system moves forward and decreases
thereby the rapidity distance between those partons and the proton remnant
system. This reduces the possibility for a large gap to occur. 

The $Q^2$ dependence is more due to the model itself.
With increasing $Q^2$ the maximum 
virtuality in the initial state parton shower increases and thereby the 
amount of radiation. Since the initial radiation tends to be along the 
incoming parton, i.e.\ the proton beam, this means more partons at rapidities
between the current and the spectator and thereby a tendency to spoil the 
gap. This gives a decrease in the relative rate of gap events. 
The exact numerical values will here
depend on the details of the parton shower which are, as discussed, not well
settled.

Although the rate of gap events in the data \cite{ZEUS,H1} is, within errors, 
essentially independent of $Q^2$ for fixed $x$, there are indications of some 
variations of the same kind as in our model. A closer comparison with the 
coming higher statistics data will therefore be interesting. 

Another testing ground for the model may be provided by data on
$F_2^D(\beta,Q^2)$ \cite{H1}. This quantity can, in pomeron-based models, be
interpreted as the pomeron $F_2$ structure function \cite{IP}, giving the
density of partons with momentum fraction $\beta$ in the pomeron. Applying
conventional QCD evolution of these partons in the pomeron should then give the
$Q^2$-dependence and explicit such calculations have been performed
\cite{IP,POMevol}. Having no pomeron in our model, one should instead consider
the evolution of partons with momentum fraction $x$ in the proton. Although
there is then no direct physical interpretation of $F_2^D$, it can still be
extracted from the model and compared with data. 
 
\ffig{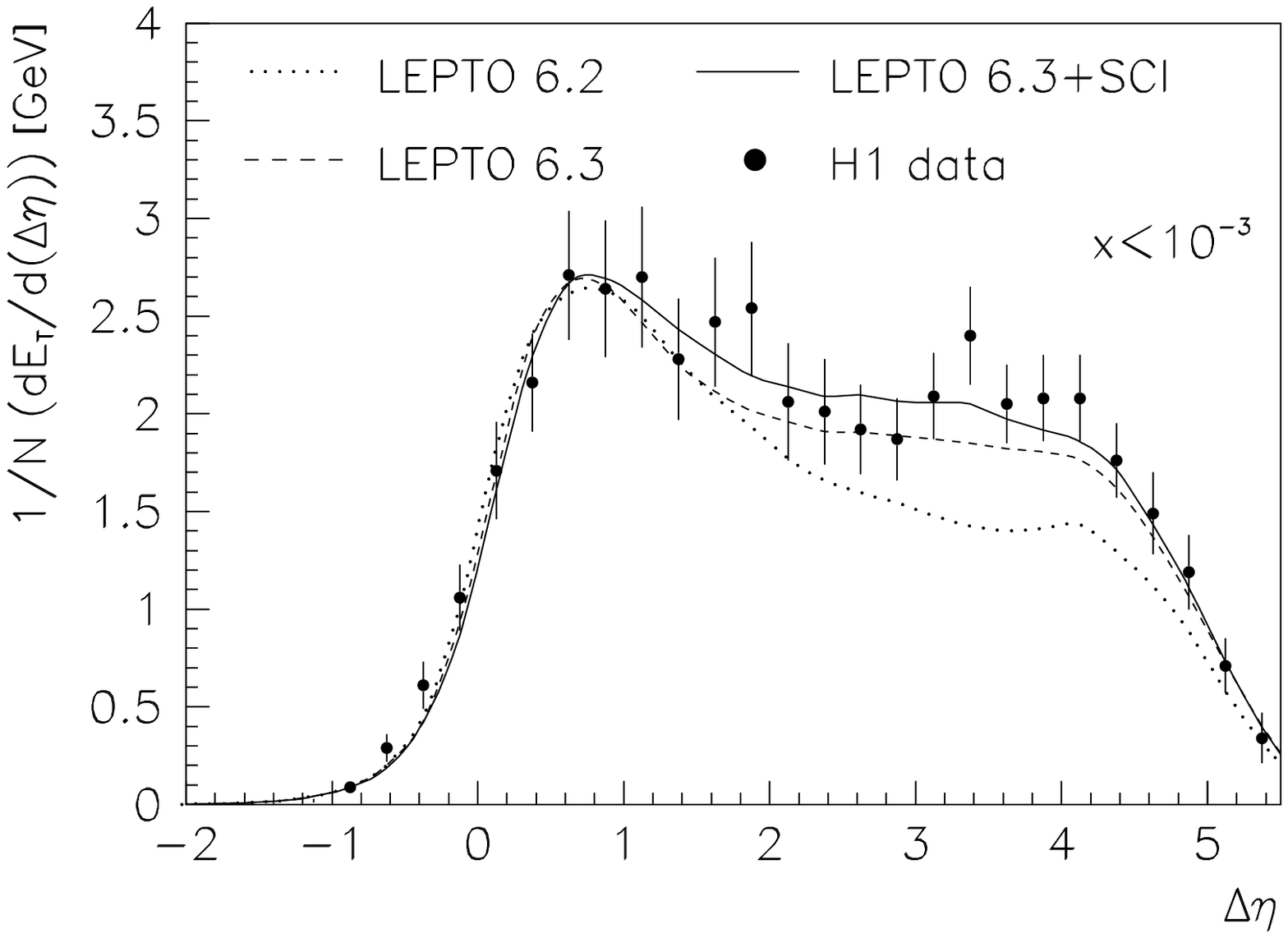}{70mm}{Transverse energy flow versus 
$\Delta\eta =\eta-\eta_{q/QPM}$, i.e.\ lab pseudorapidity relative to the 
current direction in QPM kinematics, for events with $x<10^{-3}$.  
The curves are from the earlier Monte Carlo model (6.2), 
with improved parton shower and sea quark treatment (6.3) 
and, in addition, with soft colour interactions (6.3+SCI).}{fig6}

An observable which gives complementary information relative to the rapidity 
gaps is the forward transverse energy flow. Whereas substantial initial state 
parton radiation spoils the rapidity gaps, it helps to describe the high level
of the forward energy flow. It is therefore a highly non-trivial test of any 
model that both these observables can be accounted for.
As shown in Fig.~6, the $E_T$-flow data \cite{H1eflow} can be well described
by our new model. 
The soft colour interactions generates not only gap events, but also larger
fluctuations in general. In particular, configurations may arise where 
the string goes `back and forth' and thereby produce more energy per unit 
rapidity. The effect is demonstrated by the difference between the dashed and 
full curves in Fig.~6. 

Of relevance for this observable, as well as the gap events, 
is also the modelling of the non-perturbative 
proton remnant system, i.e.\ the proton `minus' the parton entering the hard 
scattering process. In case a valence quark is removed the remainder is a 
diquark which is taken as a colour anti-triplet at the end of a string, to
which Lund model hadronization is applied. However, if a sea quark is 
removed the remainder is more complex with the valence quarks plus  
the partner anti-quark from the sea to conserve quantum numbers. 
In this study, we have improved the Monte Carlo model by assigning the 
removed quark to be a valence or sea quark and, in case of a sea quark,  
given its partner in the remnant some dynamics. 

Thus, the interacting quark is taken as a valence or sea quark from the
relative sizes of the corresponding parton distributions $q_{val}(x_1,Q^2_1)$
and $q_{sea}(x_1,Q^2_1)$, where $x_1$ is the known momentum fraction of the
quark `leaving' the proton and $Q^2_1$ is the relevant scale (typically the
cutoff $Q_0^2$ of the initial state parton shower). In case of a sea quark, the
left-over partner is given a longitudinal momentum fraction from the 
corresponding parton momentum distribution or, with similar results, from the
Altarelli-Parisi splitting function $P(g\to q\bar{q})$. In the latter case, 
the transverse momentum follows from the kinematics once the partner mass has
been fixed. This parton and the three valence quarks, which are split into a
quark and a diquark in the conventional way \cite{LEPTO}, form two colour
singlet systems (strings) together with the scattered quark (and any emitted
gluons). This two-string configuration for sea-quark initiated processes
provides a desirable continuity to the two-string gluon-induced BGF events,
thereby reducing the dependence on the value of the matrix element cut-off
$y_{cut}$. Depending on the partner sea quark momentum, the corresponding
string will extend more or less into the central region in rapidity. The
hadronization of this extra string will thereby give another contribution to
the forward energy flow, as illustrated by the difference between the dotted
and dashed lines in Fig.~6. 
 
If this new attempt to understand the rapidity gap phenomenon turns out
successful in detailed comparison with data, it will circumvent some problems 
in the pomeron-based approach. In particular those associated with the concept
of a preformed exchanged object.  
This object cannot be a real particle or state, since it has a negative 
mass square $t$. It could be a virtual exchange corresponding to 
some real state, such as a glueball, but this is presently unclear (although 
a recent glueball candidate \cite{WA91} fits on the pomeron trajectory). 
The interpretation of factorization in Regge phenomenology in terms of an
emission of a pomeron given by a pomeron flux and a pomeron-particle 
interaction cross section also has some problems. Since it is only their 
product that is experimentally observable one cannot, without further 
assumptions, define the absolute normalisation of this flux and cross section 
unambigously \cite{Landshoff}. This also gives a normalization ambiguity for 
the parton densities of the pomeron which is reflected in the problem of 
whether they fulfill a momentum sum rule or not. 
Leaving the concept of a preformed object and instead considering \cite{Soper}
a process with interactions with the proton both before and after the 
hard scattering (taking place at a short space-time scale) may avoid these 
problems, as in our model.
One may ask whether the soft colour interactions introduced here is 
essentially a model for the pomeron. This should not be the case as long as
no pomeron or Regge dynamics is introduced in the model. It remains to be seen
if the model can stand the test against data without any such extra assumptions.
In this sense, it is important to consider typical diffractive characteristics, 
e.g. the properties of the forward $R$-system, in more detailed comparisons 
with data. 

Finally, we note that our model is similar in spirit to the one of
Buchm\"uller and Hebeker \cite{BH}, which was developed independently and in
parallel with ours. Their model only considers the matrix element part, where
the whole DIS cross section is taken to be saturated by BGF, and introduces a
statistical probability that the $q\bar{q}$ pair is changed from a colour
octet into a colour singlet state. In our study, the influence of higher order
parton emissions are examined, a model for the colour exchange is introduced
and hadronization is taken into account. Having formulated our model in terms
of a Monte Carlo generator that simulates the complete final state, more
detailed comparisons with data can be made.

In conclusion, we have suggested a new mechanism for the production of events
with rapidity gaps. The basic assumption is that soft colour interactions may
occur in addition to the perturbative ones. This modifies the colour
structure for hadronisation giving colour singlet systems that are separated
in rapidity. A detailed model has features that are characteristic
for diffractive scattering and can qualitatively account for the rapidity gap
events observed in DIS at HERA.

We are grateful to W.~Buchm\"uller for interesting discussions.

\end{document}